%% file: main.tex
\documentclass[sigconf]{acmart}
\usepackage{algorithmic}
\usepackage{graphicx}
\usepackage{textcomp}
\usepackage{graphicx}
\usepackage{caption}
\usepackage{subcaption}
\usepackage{bbm}
\usepackage{color}
\usepackage{hyperref}
\usepackage{multirow}
\usepackage{enumitem}
\usepackage{algorithm} 
\usepackage{algorithmic}
\usepackage{enumerate}
\usepackage{amsmath}
\usepackage{booktabs}
\usepackage{multirow}
\usepackage{stmaryrd}
\usepackage{mathrsfs} 
\usepackage{color}
\usepackage{balance}
\usepackage{adjustbox}
\usepackage{tikz}

\usepackage[para,online,flushleft]{threeparttable}

\newcommand{\eg}{\textit{e.g.}}

\newcommand{\ie}{\textit{i.e.}}

\setlist[itemize]{leftmargin=*}
\AtBeginDocument{%
  \providecommand\BibTeX{{%
    \normalfont B\kern-0.5em{\scshape i\kern-0.25em b}\kern-0.8em\TeX}}}

\setcopyright{acmlicensed}
\copyrightyear{2018}
\acmYear{2018}
\acmDOI{XXXXXXX.XXXXXXX}

\acmConference[Conference acronym 'XX]{Make sure to enter the correct
  conference title from your rights confirmation emai}{June 03--05,
  2018}{Woodstock, NY}
\acmBooktitle{Woodstock '18: ACM Symposium on Neural Gaze Detection,
 June 03--05, 2018, Woodstock, NY} 
\acmISBN{978-1-4503-XXXX-X/18/06}

\begin{document}

\title{Beyond Positive History: Re-ranking with List-level Hybrid Feedback}

\author{Muyan Weng}
\authornote{Both authors contributed equally to this research.}
\email{1355029251@sjtu.edu.cn}
\affiliation{%
  \institution{Shanghai Jiao Tong University}
  \city{Shanghai}
  \country{China}
}

\author{Yunjia Xi}
\authornotemark[1]
\email{xiyunjia@sjtu.edu.cn}
\affiliation{%
  \institution{Shanghai Jiao Tong University}
  \city{Shanghai}
  \country{China}
}

\author{Weiwen Liu}
\email{liuweiwen8@huawei.com}
\affiliation{%
  \institution{Huawei Noah's Ark Lab}
  \city{Shenzhen}
  \country{China}
}

\author{Bo Chen}
\email{chenbo116@huawei.com}
\affiliation{%
  \institution{Huawei Noah's Ark Lab}
  \city{Shenzhen}
  \country{China}
}

\author{Jianghao Lin}
\email{chiangel@sjtu.edu.cn}
\affiliation{%
  \institution{Shanghai Jiao Tong University}
  \city{Shanghai}
  \country{China}
}

\author{Ruiming Tang}
\email{tangruiming@huawei.com}
\affiliation{%
  \institution{Huawei Noah's Ark Lab}
  \city{Shenzhen}
  \country{China}
}

\author{Weinan Zhang}
\email{wnzhang@sjtu.edu.cn}
\affiliation{%
  \institution{Shanghai Jiao Tong University}
  \city{Shanghai}
  \country{China}
}

\author{Yong Yu}
\email{yyu@sjtu.edu.cn}
\affiliation{%
  \institution{Shanghai Jiao Tong University}
  \city{Shanghai}
  \country{China}
}

\renewcommand{\shortauthors}{Yunjia Xi et al.}

\begin{abstract}
   As the last stage of recommender systems, re-ranking generates a re-ordered list that aligns with the user's preference. However, previous works generally focus on item-level positive feedback as history (\eg, only clicked items) and ignore that users provide positive or negative feedback on items in the entire list. This list-level hybrid feedback can reveal users' holistic preferences and reflect users' comparison behavior patterns manifesting within a list. Such patterns could predict user behaviors on candidate lists, thus aiding better re-ranking. Despite appealing benefits, extracting and integrating preferences and behavior patterns from list-level hybrid feedback into re-ranking multiple items remains challenging. To this end, we propose \underline{Re}-ranking with \underline{Li}st-level Hybrid \underline{Fe}edback (dubbed \textbf{RELIFE}). It captures user's preferences and behavior patterns with three modules: a Disentangled Interest Miner to disentangle the user's preferences into interests and disinterests, a Sequential Preference Mixer to learn users' entangled preferences considering the context of feedback, and a Comparison-aware Pattern Extractor to capture user's behavior patterns within each list. Moreover, for better integration of patterns, contrastive learning is adopted to align the behavior patterns of candidate and historical lists. Extensive experiments show that RELIFE significantly outperforms SOTA re-ranking baselines. 
\end{abstract}

\maketitle
\input{intro}

\input{related_work}
\input{method}
\input{exp}

\bibliographystyle{ACM-Reference-Format}
\bibliography{sample-base}

\balance
\input{appendix}

\end{document}

%% file: intro.tex
\section{Introduction}
Due to the computational limits facing numerous users and items, multi-stage recommender systems (MRS) have become ubiquitous in today's online web platforms, such as Google~\cite{seq2slate}, YouTube~\cite{wilhelm2018practical}, and Taobao~\cite{prm}. A typical MRS comprises three stages, \ie, recall, ranking, and re-ranking, with each stage narrowing down the relevant items with a more complex but more accurate model to lower response
latency \cite{liu2022neural}. 
As the final stage, re-ranking emphasizes modeling user preferences and listwise context of candidate items to generate a re-ordered list that aligns with the user's preferences. The re-ranked list is presented to the user and impacts the user's satisfaction, showcasing the crucial position of re-ranking in MRS. 

\begin{figure}[]
    \centering
    \includegraphics[clip,width=\columnwidth]{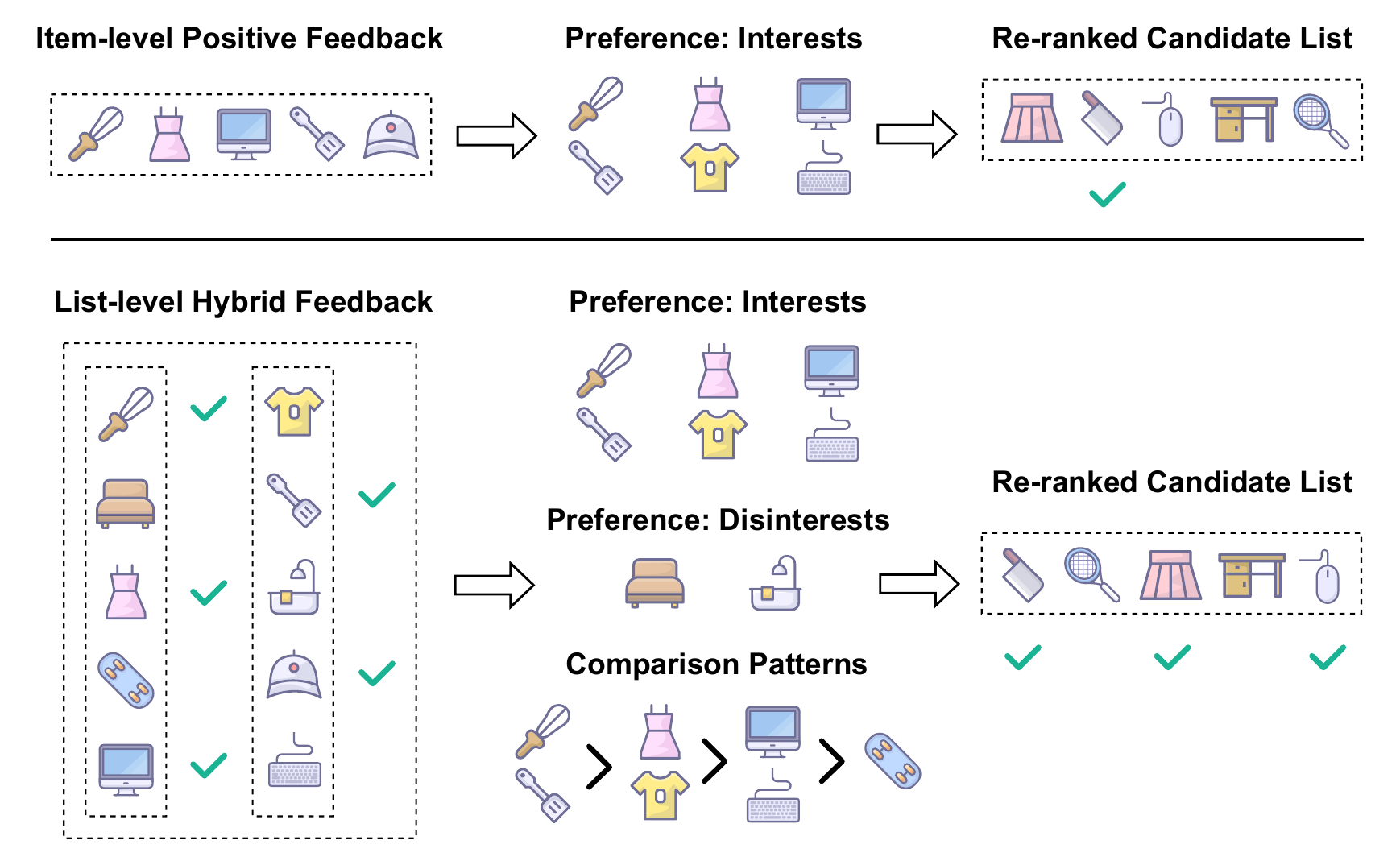}
    \vspace{-15pt}
    \caption{Comparison between re-ranking with positive feedback (above) and list-level hybrid feedback (below).}
    \vspace{-15pt}
    \label{fig:intro}
\end{figure}
Recently, much attention has been paid to re-ranking, evolving from early efforts that primarily model the listwise context of candidate lists~\cite{dlcm,prm,setrank,seq2slate,qian2022scope,irgpr} to those that emphasize user preferences and how the preferences interact with candidate items~\cite{li2022pear,xi2022multi,shi2023pier,ren2023slate}. 
Recent advancements, \eg,  MIR~\cite{xi2022multi} and PEAR~\cite{li2022pear}, significantly improve re-ranking accuracy via incorporating user preferences from historical behaviors. 
However, they leverage only users' \textbf{item-level positive feedback} as history (\eg, only clicked items), overlooking that positive feedback occurs in a specific list with negative feedback (\eg, non-clicked items) as context. In this work, we highlight that re-ranking should incorporate the user's \textbf{list-level hybrid feedback} encompassing the user's both positive and negative feedback on a list:

\textit{Firstly, list-level hybrid feedback provides more comprehensive insights into user preferences by considering both positive and negative feedback, as well as the context in which feedback occurs.} As shown in Figure~\ref{fig:intro}, list-level hybrid feedback can not only reveal users' interests and disinterests but also provide a clear context for positive feedback, \eg, surrounding negative feedback. This can mitigate biases from relying solely on positive feedback, such as exposure bias and accidental clicks. For example, if a user's past clicks are mostly on low-priced items, one might conclude that the user prefers low-priced items. However, if her negative feedback is also on low-priced items, the user may not prefer low-priced items but has been receiving such recommendations recently.
\textit{Secondly, list-level hybrid feedback can reflect the user's comparison behavior patterns beneficial for re-ranking.} Many studies have revealed users' comparison behaviors, clicking on a certain item after comparing it with surrounding ones~\cite{zhang2021constructing,fu2023f}. These comparison patterns can reveal users' preferences for different interests and be extended to the upcoming candidate lists for re-ranking. 
For instance, the user in Figure~\ref{fig:intro} is interested in kitchenware, clothing, and electronics. However, when facing kitchenware and clothing, she clicks on kitchenware, while when facing clothing and electronics, she prefers clothing. Thus, during re-ranking, we can leverage this pattern and avoid placing together items the user has strong interests in, such as kitchenware and electronics, to increase utility. 

Despite the advantages of integrating list-level hybrid feedback, its application in re-ranking presents significant challenges. \textit{First, capturing nuanced preferences that incorporate context and user's comparison behaviors is non-trivial.} Some work in CTR and sequential recommendation involves positive and negative feedback and treats user history as separate feedback sequences~\cite{xie2021deep,deng2022improving,fan2022modeling,pan2023understanding}. While they effectively glean user interests and disinterests, they neglect the context in which feedback happens and users' comparison behavior patterns in each list. The context of feedback is diverse and can change over time. Furthermore, 
users' comparative behavior patterns are influenced by the context of lists. A single list can only reflect a portion of these behavior patterns, and combining multiple lists to obtain a complete and personalized behavior pattern is challenging.
\textit{Second, how to apply these preferences and comparison behavior patterns to re-ranking is also critical.} Unlike CTR tasks considering only a single target item, re-ranking must consider items within a candidate list, emphasizing listwise context~\cite{prm,dlcm,setrank}. The dynamic interplay between historical and candidate items also enhances re-ranking~\cite{xi2022multi,xi2023bird}.  Consequently, unlike the CTR task, we must account for dynamic interactions between list-level history and candidate list. Moreover, to better incorporate comparison behavior patterns into re-ranking, it is vital to ensure that users'  behaviors on candidate lists closely align with their historical behaviors. This entails minimizing the divergence between historical and candidate behavior patterns.

Therefore, we propose a framework named \underline{Re}-ranking with \underline{Li}st-level Hybrid \underline{Fe}edback (dubbed \textbf{RELIFE}). \textit{Firstly}, to extract more comprehensive user preferences, we devise two interest extraction modules from the perspectives of disentanglement and entanglement. We design  \textbf{Disentangled Interest Miner} (DIM) to disentangle the user's preferences into interests and disinterests and introduce the dynamic interaction between history and candidate lists via co-attention. Subsequently, we develop \textbf{Sequential Preference Mixer} (SPM), which adeptly learns users' entangled preferences considering the context of feedback, temporal signals, and candidate information. \textit{Secondly}, we design \textbf{Comparaison-aware Pattern Extractor} (CPE) to capture users' comparison behavior patterns. Moreover, to better incorporate patterns into re-ranking, contrastive learning is leveraged to align behavior patterns of historical and candidate lists. Our contributions are as follows:

\begin{itemize}
    \item We emphasize the pivotal role of list-level hybrid feedback for re-ranking. To the best of our knowledge, this is the first work to incorporate the user preferences and comparison behavior patterns reflected in list-level hybrid feedback into re-ranking. 
    \item We propose RELIFE and devise several modules to exploit list-level hybrid feedback, with DIM and SPM to capture more heuristic user preferences from the perspectives of disentanglement and entanglement, and CPE and contrastive learning to extract and align the user's behavior patterns across lists.
    \item Extensive experiments on two datasets demonstrate that our method significantly outperforms state-of-the-art re-ranking algorithms with a similar level of efficiency.
\end{itemize}

%% file: related_work.tex
\section{Related Work}
Typically, multi-stage recommendation systems (MRS) are divided into three phases: recall, ranking, and re-ranking~\cite{liu2022neural}. 
The re-ranking phase reorders items by considering the user's preferences and the interrelationships among candidates. Compared to the previous two stages, re-ranking involves fewer inputs and allows for more complex models to capture user historical preferences and candidate context. It also encompasses more diverse objectives, such as utility~\cite{xi2024utility,prm,dlcm,xi2022multi,xi2023bird,xi2023device}, diversity~\cite{liu2023personalized,carraro2024enhancing,xu2023multi,lin2022feature,wu2019pd}, and fairness~\cite{han2023fair,xu2023p,naghiaei2022cpfair,peng2023re,jiang2024item}. This work focuses on utility-oriented re-ranking, aiming to optimize the overall utility of candidate list.

Early work in utility-oriented re-ranking primarily focused on modeling the mutual influence among items within the candidate list, exploring various techniques such as DNNs, RNNs, Transformers, and GNNs~\cite{ai2019learning,dlcm,seq2slate,xi2024utility}. For instance, DLCM~\cite{dlcm} encodes the whole ranking list into the representation of items via GRU. PRM~\cite{prm} adopts self-attention to encode the mutual influences between items. SRGA~\cite{qian2022scope} leverages gated attention to model the unidirectivity and locality
re-ranking. The GNN-based model IRGPR~\cite{irgpr} explicitly models relationships between items by aggregating relational information from neighborhoods on the graph. 

As re-ranking evolved, leveraging user behavior to mine user interests for personalized re-ranking also became an important direction. PEAR~\cite{li2022pear}, DIR~\cite{xi2023device}, and PFRN~\cite{huang2020personalized} utilize multi-head self-attention to capture the user's preference in historical behaviors. PEIR~\cite{shi2023pier} treats the history lists as permutations to select candidate permutations. 
MIR~\cite{xi2022multi} devises the SLAttention to mine the interactions between history list and candidate set. The above models reveal that subsuming user behaviors can significantly enhance re-ranking performance. Still, most of them focus primarily on item-level positive feedback, neglecting that list-level hybrid feedback encompassing both positive and negative feedback can provide more insights into user preferences and behavior patterns.

With the advancement of large language models (LLM), recent years have seen the emergence of applying LLM for recommendation~\cite{lin2024can,xi2024memocrs,lin2024clickprompt,xi2024decoding,xi2023towards,lin2024rella,xi2024efficient}, and some work has touched re-ranking~\cite{ma2023zero,baldelli2024twolar,pradeep2023rankvicuna,}. However, most of these have concentrated on search domains, which require substantial text and are orthogonal to re-ranking in recommendation. Therefore, we do not consider this line of work in our paper. Outside the re-ranking domain, some works extract user interests with positive and negative feedback~\cite{wang2023learning,xie2021deep,deng2022improving,fan2022modeling,pan2023understanding}. 
However, these works focus on recall or ranking stage with a single candidate item, without addressing how to integrate preferences with multiple candidate items in re-ranking.

Our work addresses this gap by considering both user preferences and behavior patterns from list-level hybrid feedback during the re-ranking process. We have designed modules to effectively integrate the extraction of user preferences and behavior patterns into the candidate list, leading to better re-ranking performance.

%% file: method.tex
\vspace{-5pt}
\section{Preliminaries}\label{sect:preli}
A re-ranking model re-orders the ranking list from ranking stage based on user preferences and interactions between candidate items. For a user $u$ with history $H_u$ and a candidate list $R_u$, the goal of re-ranking is to rearrange items in $R_u$ to optimize the overall utility of the list and output a re-ranked list $S_u$. Generally, $R_u=[x_i]^M_{i=0}$ is a list of length $M$, with each $x_i$ represents a candidate item.

In previous work, the user’s interaction history $H_u$ has mostly been represented as a list of items the user clicked on, \ie, positive feedback. In this work, we extend the history to multiple lists, represented as $H_u=[h_i]^N_{i=1}$ of length $N$, where each $h_i$ is a historical list interacted by the user that includes both clicked (positive feedback) and non-clicked (negative feedback) items of length $M$. Thus, the entire user history can be formulated as an item matrix $H_{N\times M}$, where each entry $h_{i,j}$  represents the $j$-th item in the $i$-th history list. Correspondingly, we use a feedback matrix $F_{N \times M}$, where each entry $f_{i,j}\in\{0,1\}$ indicates the type of feedback the user gave to the item $h_{i,j}$, with $f_{i,j}=1$ for positive feedback and $f_{i,j}=0$ for negative feedback. This extension of user interaction history allows for more comprehensive modeling of user preferences and behaviors, thereby further optimizing the utility of the re-ranked results.

\section{Methodology}

\begin{figure*}
    \centering
    \vspace{-10pt}
    \includegraphics[clip,width=\textwidth]{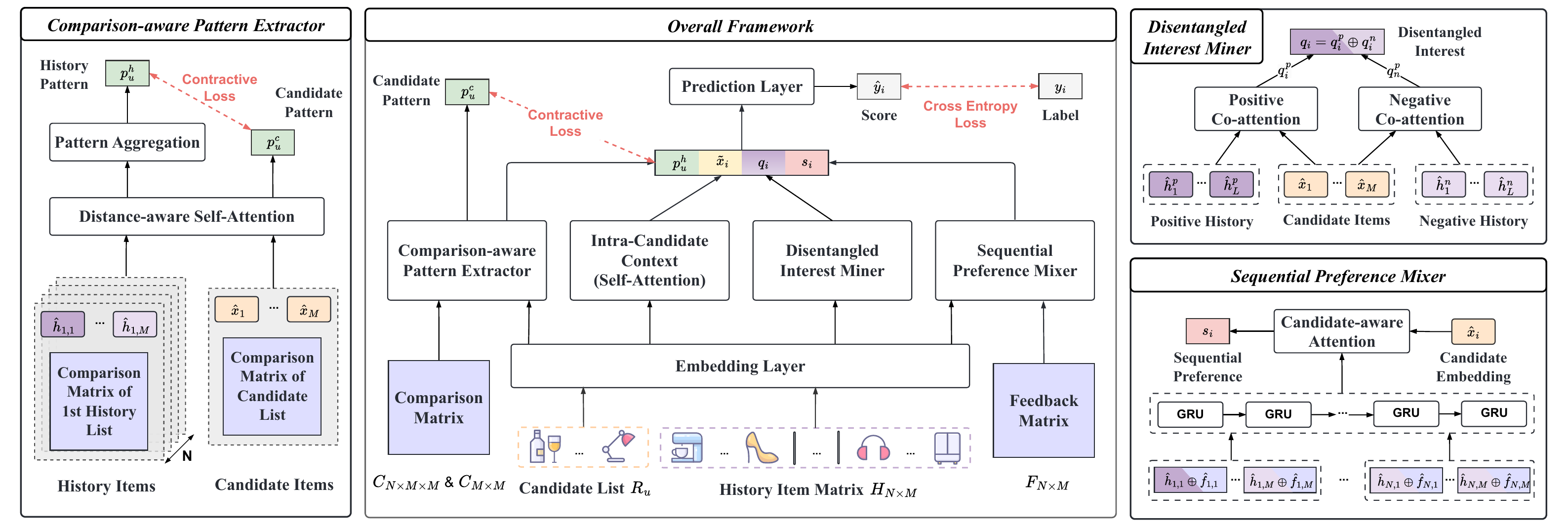}
    \vspace{-15pt}
    \caption{The overall framework of RELIFE mainly consists of Intra-Candidate Context (ICC), Disentangled Interest Miner (DIM), Sequential Preference Mixer (SPM), and Comparison-aware Pattern Extractor (CPE).}
    \vspace{-10pt}
    \label{fig:framework}
\end{figure*}

\subsection{Overview}

The framework of RELIFE, as shown in Figure~\ref{fig:framework}, consists of the following parts: embedding layer, Intra-Candidate Context (ICC), Disentangled Interest Miner (DIM), Sequential Preference Mixer (SPM), Comparison-aware Pattern Extractor (CPE), and prediction layer. First, the embedding layer converts sparse categorical features of items into dense embeddings. The Intra-Candidate Context (ICC) modeling captures the interactions within the candidate list through self-attention. Then, the Disentangled Interest Miner (DIM) utilizes two co-attention~\cite{lu2016hierarchical} modules to disentangle the user preference towards candidate items into interests and disinterests. Next, the Sequential Preference Mixer (SPM) learns the user's entangled preference towards each candidate item, considering both positive and negative feedback across multiple historical lists and temporal information. Subsequently, the Comparison-aware Pattern Extractor (CPE) obtains user patterns within each list, mainly concerning comparison behaviors. Moreover, we employ contrastive learning to align the user's historical patterns and the pattern of candidate list. Lastly, the prediction layer outputs scores for candidate items.

\subsection{Embedding \& Intra-Candidate Context (ICC)}
Our model takes as input the candidate list $R_u$ and historical item matrix $H_{N\times M}$ from multiple history lists. These items in candidate and historical lists and their associated features predominantly constitute highly sparse multi-domain categorical features. Consequently, an embedding layer is required to convert these items and their features from sparse raw representations into dense vectors. Specifically, for an item $x_i$ within the candidate list $R_u$, we employ a projection matrix to transform the sparse features related to $x_i$, such as item ID and category ID, into dense vectors. These dense vectors are subsequently concatenated to form $\hat{x}_i\in\mathbb{R}^{d_x}$ with a dimensionality of $d_x$, thus the entire set of candidate items to be represented as $\hat{\mathbf{X}}\in\mathbb{R}^{M\times d_x}$. Similarly, each item $h_{i,j}$ in the user's historical item matrix is transmuted into $\hat{h}_{i,j}\in\mathbb{R}^{d_h}$ of dimension $d_h$ via the projection matrix, yielding the historical feature matrix $\hat{\mathbf{H}}\in\mathbb{R}^{N\times M\times d_h}$ for the user’s multi-list behavior history. The embedding layer for historical items and candidate items is shared. Moreover, beyond items, the user's feedback is also discrete (\ie, 0 or 1) and necessitates conversion into dense features. Thus, each feedback $f_{i,j}$ of feedback matrix $F\in\mathbb{R}^{N\times M}$ is converted into $\hat{f}_{i,j}\in\mathbb{R}^{d_f}$ of size $d_f$ through another projection matrix of dimension $2\times d_f$, resulting in the feedback feature matrix $\hat{\mathbf{F}}\in\mathbb{R}^{N\times M\times d_f}$.

As validated in previous work~\cite{prm,dlcm,xi2022multi}, a pivotal factor affecting re-ranking is the interaction among the items within the candidate list, referred to as the listwise context. Drawing from prior research~\cite{prm,xi2022multi,li2022pear}, we employ self-attention~\cite{vaswani2017attention} to capture the mutual influence between any two candidate items. Taking the candidate item feature matrix $\hat{\mathbf{X}}$ derived from the embedding layer as input, we can obtain the interrelated candidate items through the following common attention operations:
\begin{equation}
     \Tilde{\mathbf{X}}=\text{Attention}(\hat{\mathbf{X}})=\text{softmax}\left(\frac{(\hat{\mathbf{X}}\mathbf{W}_Q)(\hat{\mathbf{X}}\mathbf{W}_K)^\top}{\sqrt{d_a}}\right)(\hat{\mathbf{X}}\mathbf{W}_V)\,,
     \label{eq:self-attention}
\end{equation}
where the output $\Tilde{\mathbf{X}}\in\mathbb{R}^{M\times d_x}$ is the attended matrix. In terms of implementation, we apply the multi-head mechanism to increase the stability of the self-attention network. Finally, we get the interacted item representation $\Tilde{x}_i\in\mathbb{R}^{d_x}$ for the $i$-th item in the candidate list $R_u$, that is the $i$-th row of the attended matrix $\Tilde{\mathbf{X}}$.

\subsection{Disentangled Interest Miner (DIM)}
Users exhibit their preferences, including interests and disinterests, towards items displayed in the current list, and these attitudes are reflected in their historical behaviors through positive and negative feedback. The dynamic interaction between historical feedback and current candidate items may also influence these interests and aversions~\cite{xi2022multi}. For example, different historical items may contribute variably to the re-ranking of different candidate lists. Therefore, we design DIM to disentangle the user's preferences into interests and disinterests that are useful for the candidate items.  We first separate the user's lits-level history into positive and negative parts, and then, motivated by~\cite{xi2022multi}, co-attention~\cite{lu2016hierarchical} is adapted to fuse information from both positive/negative history and candidate, extracting useful information for re-ranking. 

As this module only considers the user's preference with positive/negative feedback, we flatten the user's multi-list historical behaviors into two sequences based on the type of feedback, thus dividing the historical feature matrix $\hat{\mathbf{H}}$ into two matrices: the positive interaction matrix $\hat{\mathbf{H}}^p\in\mathbb{R}^{L\times d^f}=[\hat{h}^p_i]^L_{i=1}$ and the negative interaction matrix $\hat{\mathbf{H}}^n\in\mathbb{R}^{L\times d^f}=[\hat{h}^n_i]^L_{i=1}$. Here, $\hat{h}^p_i$
  and $\hat{h}^n_i$ represent the item embeddings with positive and negative feedback, respectively, and we pad the sequences to the maximum length $L$. The DIM employs two co-attentions to separately mine preferences from positive and negative feedback that are beneficial for the candidate set. For positive feedback, for instance, the co-attention takes information from both historical and candidate sources, \ie, $\hat{\mathbf{H}}^p$  and $\hat{\mathbf{X}}$, and attends to the two sources simultaneously by calculating the similarity between any pairs of items in history and candidate. This similarity, \ie, affinity matrix $\mathbf{E}\in\mathbb{R}^{N\times L}$, is computed as follows,
  \begin{equation}
      \mathbf{E}^p=\tanh{(\hat{\mathbf{X}}\mathbf{W}_{e}^p(\hat{\mathbf{H}}^p)^T)}\,,
      \label{eq:affinty}
  \end{equation}
  where the learnable matrix $\mathbf{W}_e^p\in\mathbb{R}^{d_x\times d_h}$ is the importance of the association between any pair of items in history and the candidate list. Next, the affinity matrix $\mathbf{E}^p$, which integrates information from both history and candidates, is utilized to predict the attention weights via the following process: 
    \begin{equation}
     \begin{split}
     \mathbf{A}_x^p&=\text{softmax}(\tanh{(\hat{\mathbf{X}}\mathbf{W}_x^p+\mathbf{E}^p(\hat{\mathbf{H}}^p\mathbf{W}_h^p))})\, ,\\
    \mathbf{A}_h^p&=\text{softmax}(\tanh{((\hat{\mathbf{H}}^p\mathbf{W}_h^p)^T+\hat{\mathbf{X}}\mathbf{W}_x^p\mathbf{E}^p)}). \\
    \end{split}
     \label{eq:co-attn}
 \end{equation}
 where $\mathbf{W}_x^p\in\mathbb{R}^{d_x\times M },\mathbf{W}_h^p\in\mathbb{R}^{d_h\times M}$ are learnable weight matrices. Attention weight matrices $\mathbf{A}_x^p\in\mathbb{R}^{M\times M}$ and $\mathbf{A}_h^p\in\mathbb{R}^{M\times L}$ are the attention probabilities of items in the candidate and history list, respectively, which preserve useful information for re-ranking. With $\mathbf{A}_x^p$ and $\mathbf{A}_h^p$, attention vectors for candidate and history items are acquired as weighted sum of items in candidate and history list, \ie, 
 \begin{equation}
    \Tilde{\mathbf{X}}^p=\mathbf{A}_x^p\hat{\mathbf{X}}\, ,\,\,\,
    \Tilde{\mathbf{H}}^p=\mathbf{A}_h^p\hat{\mathbf{H}}^p.
    \label{eq:co-attn-out}
 \end{equation}
  where $\Tilde{\mathbf{X}}^p=[\Tilde{x}^p_i]^M_{i=1}$ and $\Tilde{\mathbf{H}}^p=[\Tilde{h}^p_i]^M_{i=1}$ are the interacted representation matrices containing useful information from both candidate and positive history list. Therefore, we concatenate representations from both candidate and history as the user's positive interest for each candidate item $x_i$, \ie, $q_i^p=\Tilde{x}^p_i\oplus\Tilde{h}^p_i, i=1,\ldots,M$, where $\oplus$ denotes the concatenation operation. 
  
  Similarly, after applying co-attention to candidate matrix $\hat{\mathbf{X}}$ and negative history matrix $\hat{\mathbf{H}}^n$, we can obtain the interacted representation matrices $\Tilde{\mathbf{X}}^n=[\Tilde{x}^n_i]^M_{i=1}$ and $\Tilde{\mathbf{H}}^n=[\Tilde{h}^n_i]^M_{i=1}$. Then, the user's negative interest for each item is calculated by $q_i^n=\Tilde{x}^n_i\oplus\Tilde{h}^n_i, i=1,\ldots,M$. Finally, the \textit{disentangled interest} representation $q_i$ is a splice of the user's positive and negative interests for each candidate item, that is, $q_i = q^p_i\oplus q^n_i, i=1,\ldots,M$.

\subsection{Sequential Preference Mixer (SPM)}
Previous studies have predominantly leveraged positive feedback to infer user preferences from historical behaviors. However, this approach can introduce bias since it neglects the context of positive feedback created by abundant negative feedback. This negative feedback can provide critical insights that positive feedback alone cannot capture. For instance, if a user's historical clicks are centered on low-priced items, relying solely on positive feedback might lead us to conclude that the user prefers low-priced products. However, if the negative feedback also consists of low-priced items, it may indicate that the user might not actually prefer low-priced products; instead, the recent recommendations from the system were predominantly low-priced items. Hence, it is imperative to model the interaction between positive and negative feedback to glean a more holistic understanding of user preferences.

Besides, user preferences evolve over time, so we incorporate temporal information into the interaction modeling between positive and negative feedback in SPM. First, we chronologically arrange the positive and negative feedback from multiple lists into a single sequence, and then a GRU is employed to preserve the sequential dependencies. Subsequently, candidate-aware attention is employed on the historical sequence to capture the user's holistic preferences relevant to each candidate item.

In particular,  we first place the items of the historical feature matrix $\hat{\mathbf{H}}$ in chronological order to obtain a sequence of length $NM$, denoted as $\mathbf{H}^s=[h^s_i]^{NM}_{i=1}$. Similarly, we process the feedback feature matrix $\hat{\mathbf{F}}$
 in the same way to obtain the feedback sequence $\mathbf{F}^s=[f^s_i]^{NM}_{i=1}$. Next, we concatenate the features of each item with the corresponding feedback to obtain the input for the next module $\hat{\mathbf{H}}^s=[\hat{h}^s_i]^{NM}_{i=1}$, where $\hat{h}^s_i=[h^s_i\oplus f^s_i], i=1,\ldots,NM$ and $\oplus$ denotes concatenation operation. To retain the temporal patterns in the user's historical behavior, we leverage a GRU~\cite{cho-etal-2014-properties} to model the user's evolving interests as follows
  \begin{equation}
    \Tilde{\mathbf{H}}^s = \text{GRU}(\hat{\mathbf{H}}^s),
    \label{eq:gru}
 \end{equation}
 where $\Tilde{\mathbf{H}}^s=[\Tilde{h}^s_i]^{NM}_{i=1}$ represents GRU's output and $\Tilde{h}^s_i,i=1,\ldots,NM$ is the interacted representation for the $i$-th item in sequence $\hat{\mathbf{H}}^s$.

Subsequently, candidate-aware attention is employed to extract the user's fine-grained preferences relevant to the candidate items from the historical sequence, which now carries both temporal and candidate information. For each candidate item $\hat{x}_i$ in the candidate list $R_u$, we compute its similarity with each historical item to obtain attention coefficients. These coefficients are utilized to perform a weighted sum of the historical items. This process is as follows,
\begin{equation}
    s_i = \sum_{j=1}^{NM}w_j\Tilde{h}^s_j=\sum_{j=1}^{NM}a(\hat{x}_i,\Tilde{h}^s_j)\Tilde{h}^s_j,
    \label{eq:feedback-attn}
 \end{equation}
 where $a(\cdot)$ is a feed-forward
network with softmax, calculating the similarity between each historical item $\Tilde{h}^s_j$ and the target candidate item $\hat{x}_i$, and the output $s_i$ denotes the user's \textit{sequential preference} towards candidate item $\hat{x}_i$.

\subsection{Comparison-aware Pattern Extractor (CPE)}
The user's positive and negative feedback on a list is not only determined by their interests but sometimes also influenced by their behavioral habits, particularly their comparison behaviors. Many studies have pointed out that users often compare an item with surrounding items before clicking on it~\cite{zhang2021constructing,fu2023f}. This suggests that the user's feedback on a target item is also related to its nearby items. For example, if a user is currently interested in dresses and the list contains only dresses, she might compare the images of these dresses and choose the one that best matches her preferences. However, if the list contains only one dress, the user is likely to click on it even if it is not her favorite, simply because the surrounding items do not match her desires. Furthermore, due to the continuity of user behaviors, the patterns displayed in historical lists are likely to manifest in the upcoming candidate lists for re-ranking. Aligning user behavior patterns between historical and candidate lists can help us better distinguish user interests in candidate items. Therefore, when re-ranking a candidate list, it is essential to consider the user's comparison behavior patterns embodied in historical lists.

To address this, we design CPE based on the distance-aware self-attention mechanism, which adjusts the attention weights according to the distance between positive and negative feedback. Consequently, the modeling of positive feedback is more influenced by the nearby negative feedback, making it more reflective of the user's comparison behavior. Next, we leverage CPE to extract behavior patterns from historical and candidate lists, subsequently utilizing contrastive learning to align these patterns. This approach ensures the alignment of user behavior between historical and candidate lists, thereby facilitating more effective re-ranking.  

First, we construct a comparison matrix for historical lists to adjust the attention weight. This matrix is based on the distance between the clicked item and surrounding items, allowing positive feedback items to better account for the influence of nearby negative feedback. Since we have $N$ historical lists, each containing 
$M$ items, we will construct a matrix 
$C\in\mathbb{R}^{N\times M\times M}=[C_t]^N_{t=1}$, where $C_t$ representing the comparison matrix of the $t$-th historical list, and it will adjust the attention weight within the 
$t$-th historical list. The construction of each entry $c_{t,i,j}$ in the matrix $C_t$ is expressed as:
 \begin{equation}
 c_{t,i,j} = 
\begin{cases} 
|i - j|, & f_{t,i} = 1 \\
0, & \text{otherwise}
\end{cases}
\label{eq:compar-mat}
 \end{equation}
 where $f_{t,i}$ is the feedback for the $i$-th item in the $t$-th historical list.
Considering that closer negative feedback should have a greater impact on positive feedback, we use a learnable sigmoid function~\cite{wu2021transformer} as a scaling function to transform the distance matrix, thereby making the influence inversely proportional to the distance. The learnable sigmoid function $f$ maps each entry $c_{t, i, j}$ of comparison matrix $C$ to a positive influence factor $\hat{c}_{t,i,j}$ of range $(0,1]$, $f(\cdot|v):\mathbb R^*\rightarrow \mathbb (0,1]$.
\begin{equation}
    \hat{c}_{t,i,j}=f(c_{t,i,j}|v)=\frac{1+\exp(v)}{1+\exp(v + \sigma c_{t,i,j})}\,,
\end{equation}
where a learnable scalar $v\in\mathbb R$ determines the steepness of function $f(\cdot|v)$, and a hyper-parameter $\sigma > 0$ normalizes the distance $c_{t,i,j}$ and stabilizes the training.  Note that $f(\cdot|v)$ is a monotonically decreasing function w.r.t. the input and satisfies $f(0|v)=1$ and $f(+\infty|v)\rightarrow 0$, so that the influence of between the items gradually decreases from $1$ to $0$ as the input distance grows.

Next, the influence factor $\hat{c}_{t,i,j}$ is employed to scale the pairwise mutual influence between items on the historical list. Self-attention is adopted for modeling these interactions, while the attention weights are scaled according to the influence factor. Suppose $\hat{\mathbf{H}}_t\in\mathbb{R}^{M\times d_h}$ is the item feature matrix for the $t$-th historical list. We form all the $\hat{c}_{t,i,j}$ for the $t$-th historical list into a matrix $\hat{C}_{t}$, and numerically scale the preliminary self-attention weights:
\begin{equation*}
\begin{split}
    \mathbf{O}_t&=\operatorname{Softmax}\left(\frac{\phi((\hat{\mathbf{H}}_t\mathbf{W}_Q)(\hat{\mathbf{H}}_t\mathbf{W}_K)^\top)\odot \hat{C}_t}{\sqrt{d_a}}\right)(\hat{\mathbf{H}}_t\mathbf{W}_V),
\end{split}
\end{equation*}
where $\odot$ is the element-wise product, the preliminary attention weights $(\hat{\mathbf{H}}_t\mathbf{W}_Q)(\hat{\mathbf{H}}_t\mathbf{W}_K)^\top)$ are adjusted by the influence factors. Note that $\phi(\cdot)$ is a non-negative monotonically increasing function introduced to avoid negative attention weights, as negative preliminary attention weights can invert the positive influence factor and violate the negative correlation between distances and influences. Here, we use the $\operatorname{softplus}(\cdot)$ function as $\phi(\cdot)$. In terms of implementation, we apply the multi-head mechanism to increase the stability of the self-attention network. The output $\mathbf{O}_t=[o_i]^M_{i=1}$ consists of the interacted representation of $m$ items in the $t$-th historical list. We average these to obtain the pattern representation $p_t = \frac{1}{M}\sum^M_{t=1} o_i$ for the $t$-th historical list. 
  
Then, we devise pattern aggregation to aggregate the patterns obtained from the $N$ historical lists to derive the intra-list history pattern. Since different lists may contribute differently to the final historical pattern, we aggregate $p_t,t=1,\ldots,N$ with an attention mechanism \cite{yang-etal-2016-hierarchical} to form the \textit{intra-list history pattern} $p^h_u$ for user $u$:
\begin{equation}
\begin{split}
    g_{t} &= \tanh{(\mathbf{W}_lp_t + b_l)}\,,\\
    \alpha_{t}& =\frac{\text{exp}(g_{t}^\top \cdot e^h)}{\sum_{j=1}^{N} \text{exp}(g_{t}^\top\cdot e^h)}\,,\\
    p^h_u& =\sum\nolimits_{t=1}^N{\alpha_{t}p_t}\,,\\
\end{split}
\end{equation}
 where $\mathbf{W}_l\in\mathbb R^{d_h\times d_h}$ and $b_l\in\mathbb R^{d_h}$ are the learnable weights. The importance of each pattern is measured by the similarity of $g_{t}$ with an item-level query vector $e^{p}\in\mathbb{R}^{d_h}$, a trainable parameter. Next, we normalize the weights $\alpha_{t}$ and compute the history pattern $p^h_u$ for the user $u$ by the weighted sum of each item. This pattern is an aggregation of user behaviors across multiple historical lists, which is not only incorporated into the prediction of candidate items but is also used in the following contrastive learning to align user behaviors between historical and candidate lists.

Next, to obtain the behavior pattern on the candidate list, we apply CPM similarly on the candidate list to obtain the \textit{candidate pattern} $p^c_u$ for user $u$, with click labels of candidate list $R_u$ as the feedback. The only difference lies in that, unlike the historical lists, there is only one candidate list, so there is no need for pattern aggregation. Once we have the history and candidate patterns, contrastive learning can be applied to align the two patterns of the same user. Here, we employ the InfoNCE loss~\cite{oord2018representation} as the contrastive learning training objective:
\begin{equation}
\mathcal{L}_{info} = -\frac{1}{|\mathcal{U}|}\sum_{u\in\mathcal{U}}\log \frac{\exp(p_u^c\cdot p^h_u/\tau)}{\exp(\sum_{u'\in\mathcal{U}}p_{u'}^c\cdot p^h_{u}/\tau)}.
\label{eq:infoNCE}
\end{equation}

Note that the candidates' labels are incorporated for candidate patterns; however, this is solely utilized in training to optimize InfoNCE loss. During inference, candidate patterns are not employed, thereby eliminating the risk of information leakage.

\subsection{Prediction and Optimization}
Finally, we adopt a prediction layer, a multi-layer perceptron (MLP), to integrate the representations extracted by the previous modules and produce the predicted scores for each candidate item. Its input comprises the disentangled interest $q_i$, history pattern $p^h_u$, sequential preference $s_i$, and context-modeled item representation $\Tilde{x}_i$ that we previously obtained. The output predicted score $\hat{y}_i$ for the $i$-th item in the ranking list $R_u$ is computed as follows,
\begin{equation}
    \hat{y}_{i} = \text{MLP}([q_i\oplus p^h_u\oplus s_{i}\oplus \Tilde{x}_i])\,,
    \label{eq:final_mlp}
\end{equation}
where the LeakyReLU activate function is applied in MLP, and $\oplus$ represents the concatenation operation. The final rankings, thus, can be achieved by sorting the items
by their predicted scores.

The optimization function for our model can be divided into two parts. One part is the utility loss, which enhances the accuracy of the re-ranking, and the other is the pattern contractive loss concerning user behavior patterns. We adopt infoNCE loss for pattern contractive loss and aim to reduce the distance between the user's historical pattern and the candidate pattern, thereby making the user's behavior on the candidate set more similar to their historical behavior.

For the utility loss, we use the common cross-entropy loss. Given the click labels 
$Y_u=[y_1, y_2, \ldots, y_M]$ for each user 
$u$ on the candidate list $R_u$, this loss is calculated as follows:
\begin{equation}
    \mathcal{L}_{util} = -\frac{1}{|\mathcal{U}|}\sum_{u\in\mathcal{U}}\sum^{M}_{i=1} y_i\log \hat{y}_{i}+(1-y_{i})\log (1-\hat{y}_{i}),
    \label{eq:ce}
\end{equation}
where $\hat{y}_{i}$ is the predicted score for the $i$-th candidate item in candidate list $R_u$, and $\mathcal{U}$ is the user set. As for the pattern contrastive loss, $\mathcal{L}_{info}$, it has been detailed in Section 4.4. Lastly, combining Eq~\eqref{eq:ce} and \eqref{eq:infoNCE}, we derive the final optimization loss as follows:
\begin{equation}
    \mathcal{L} = \mathcal{L}_{util} + \beta \mathcal{L}_{info},
    \label{eq:loss}
\end{equation}
where $\beta$ denotes the hyper-parameter that controls the tradeoff between the utility and the contrastive loss.

\subsection{Complexity Analysis}
Here, we analyze the time complexity of RELIFE and find it is comparable with SOTA baselines. Assuming $n$ and $m$ denote the number of historical lists and the number of items in each list, SPM leverages GRU and takes $O(nm)$. The ICC modeling involves self-attention of candidates, and its complexity is $O(m^2)$. The time complexity of DIM is $O(nm^2)$ as it mainly contains the co-attention between historical and candidate lists~\cite{xi2022multi}. CPE involves separate self-attention for historical and candidate lists, which takes $O((n+1)m^2)$. Overall, the time complexity of RELIFE is $O(nm^2)$.  The value of $n$ is usually small, \eg, 5, so it can be considered that RELIFE's complexity is comparable with most SOTA re-ranking models (\ie, $O(n^2m^2)$ for PIER~\cite{shi2023pier}) and $O(k^2+m^2+km))$ for MIR~\cite{xi2022multi} and  PEAR~\cite{li2022pear} where $k$ is the number of positive historical feedback. 

%% file: exp.tex
\section{Experiment}

In this section, we tend to address the following research questions (RQs) to gain more insights into RELIFE.  
\begin{itemize}
    \item \textbf{RQ1:}  How does RELIFE perform compared to other state-of-the-art re-ranking baselines?
    \item \textbf{RQ2:} What roles do RELIFE's modules play in its performance? 
    \item \textbf{RQ3:} What is the impact of some hyperparameters of RELIFE on its performance?
    \item \textbf{RQ4:} How does RELIFE's performance in terms of efficiency compare to other baselines?
    
\end{itemize}

\subsection{Setup}
\begin{table*}[]

    \vspace{-10pt}
    \caption{Overall performance on two benchmark datasets. The best result is given in bold, while the second-best value is underlined. The symbol * indicates statistically significant improvement over the best baselines( t-test with $p < 0.05$).}
    \vspace{-8pt}
    \centering
    \scalebox{1}{
    \setlength{\tabcolsep}{1.5mm}
    {
  \begin{tabular}{ccccccc|cccccc}
\toprule
\multirow{4}{*}{Model} & \multicolumn{6}{c|}{PRM Public}                                                                                               & \multicolumn{6}{c}{MovieLens-20M}\\ \cmidrule{2-13} 
                       & \multicolumn{3}{c|}{@5}                           & \multicolumn{3}{c|}{@10}                            & \multicolumn{3}{c|}{@5}                                                                        & \multicolumn{3}{c}{@10}                                                               \\ \cmidrule{2-13} 
                       & MAP             & NDCG            & \multicolumn{1}{l|}{Click}        & MAP             & NDCG            & Click         & \multicolumn{1}{l}{MAP}    & \multicolumn{1}{l}{NDCG}   & \multicolumn{1}{l|}{Click}         & \multicolumn{1}{l}{MAP}    & \multicolumn{1}{l}{NDCG}   & \multicolumn{1}{l}{Click} \\ \midrule
DLCM                   & 0.3074          & 0.2930           & \multicolumn{1}{l|}{0.6180}          & 0.3236          & 0.3456          & 0.8853          & 0.7327                     & 0.6836                     & \multicolumn{1}{c|}{2.7447}          & 0.6953                     & 0.7956                     & 3.9795                      \\
SAR                    & 0.3092          & 0.2965          & \multicolumn{1}{l|}{0.6414}         & 0.3245          & 0.3546          & 0.9362          & 0.7276                     & 0.6751                     & \multicolumn{1}{c|}{2.7176}          & 0.6879                     & 0.7911                     & 3.9562                      \\
PRM                    & 0.3173          & 0.2958          & \multicolumn{1}{l|}{0.6174}         & 0.3290           & 0.3465          & 0.8794          & 0.7373                     & 0.6896                     & \multicolumn{1}{c|}{2.7621}          & 0.7004                     & 0.7989                     & 3.9752                      \\
SetRank                & 0.3246          & 0.3055          & \multicolumn{1}{l|}{0.6442}         & 0.3360           & 0.3636          & 0.9431          & 0.7324                     & 0.6834                     & \multicolumn{1}{c|}{2.7425}          & 0.6957                     & 0.7955                     & 3.9705                      \\
SRGA                   & 0.3196          & 0.3026          & \multicolumn{1}{l|}{0.6356}         & 0.3366          & 0.3567          & 0.9069          & 0.7378 & 0.6893 & \multicolumn{1}{c|}{2.7675}          & 0.7001 & 0.7989 & 3.9884  \\ 
\midrule

DFN  &   0.3243  & 0.3083       &\multicolumn{1}{l|}{0.6531}     &0.3386    &0.3636       &0.9352         &0.7488    &0.7015   & \multicolumn{1}{c|}{2.7968}    &  0.7112  & 0.8065   & \underline{4.0022}  \\
RACP  &  0.3169   & 0.3019     & \multicolumn{1}{l|}{0.6411}         & 0.3344      & 0.3579   & 0.9250    & \underline{0.7502}    &0.7016   & \multicolumn{1}{c|}{2.7990}    & \underline{0.7115}    & \underline{0.8070}  & 4.0017  \\
\midrule
PEAR                   & 0.3133          & 0.2987          & \multicolumn{1}{l|}{0.6388}         & 0.3281          & 0.3571          & 0.9369          & 0.7446 & 0.6965 & \multicolumn{1}{c|}{2.7809}          & 0.7070 & 0.8036 & 3.9892  \\
MIR                    & \underline{0.3284}          & \underline{0.3113}         & \multicolumn{1}{l|}{0.6546}         & \underline{0.3431}          & 0.3663          & 0.9353          & 0.7473                     & 0.6987                     & \multicolumn{1}{c|}{2.7872}          & 0.7088                     & 0.8050                     & 3.9999                      \\
PIER                   & 0.3235          & 0.3096          & \multicolumn{1}{l|}{\underline{0.6586}}         & 0.3414          & \underline{0.3682}        & \underline{0.9554}          & 0.7492                     & \underline{0.7019}                     & \multicolumn{1}{c|}{\underline{2.7996}}          & 0.7113                     & 0.8067                     & 3.9994                      \\
\midrule
\textbf{RELIFE}        & \textbf{0.3393*} & \textbf{0.3225*} & \multicolumn{1}{l|}{\textbf{0.6820*}} & \textbf{0.3551*} & \textbf{0.3805*} & \textbf{0.9792*} & \textbf{0.7585*}            & \textbf{0.7117*}            & \multicolumn{1}{c|}{\textbf{2.8292*}} & \textbf{0.7202*}            & \textbf{0.8130*}             & \textbf{4.0122}             \\ \bottomrule
\end{tabular}  
    }
    }

\label{tab:overall}
\vspace{-8pt}
\end{table*}

\subsubsection{Datasets}
Our experiments are conducted on two public datasets, PRM public\footnote{\url{https://github.com/rank2rec/rerank}} and MovieLens-20M\footnote{\url{https://grouplens.org/datasets/movielens/20m/}}. \textbf{PRM Public} contains 7,246,323 items, 743,720 users, and 14,350,968 records from a real-world e-commerce recommender system. Each record is a recommendation list of 30 items and the user's feedback on them. For each user, the last recommendation list in her interacted lists is utilized as the candidate list for re-ranking, and the previous interacted lists are used as history lists. \textbf{MovieLens-20M} records 20 million ratings provided by 138, 493 users for 27,278 movies. Although it is not presented in list format, many re-ranking studies utilize this dataset to generate simulated lists for experiments\cite{ren2023slate,liu2023personalized,wu2019pd}. Due to the difficulty in finding datasets that contain both candidate lists and historical behaviors in list form, we follow the approach in re-ranking~\cite{liu2023personalized} and use the ML-20M dataset along with a dependent click model (DCM)~\cite{guo2009efficient,liu2018contextual,katariya2016dcm} to generate \textit{simulated candidate lists}. More details about datasets can be found in Appendix~\ref{sec:dataset}

\subsubsection{Baselines}  
We compare RELIFE with several SOTA re-ranking models, which are categorized into two main types. The first one includes models that emphasize the interactions between items in the candidate list, such as \textbf{DLCM}~\cite{dlcm}, \textbf{PRM}~\cite{prm}, \textbf{SetRank}~\cite{setrank}, \textbf{SRGA}~\cite{qian2022scope}, \textbf{SAR}~\cite{ren2023slate}. The second type focuses on extracting users' historical interests for re-ranking, including \textbf{PEAR}~\cite{li2022pear}, \textbf{MIR}~\cite{xi2022multi}, and \textbf{PIER}~\cite{shi2023pier}. Besides, we also compare two models in the CTR domain that utilize positive and negative feedback, \textbf{DFN}~\cite{xie2021deep} and \textbf{RACP}~\cite{fan2022modeling}. More detailed descriptions can be found in Appendix~\ref{sec:baseline}.

\subsubsection{Evaluation Metrics } 
Our proposed model and baselines are evaluated using both ranking and utility metrics ~\cite{xi2024utility,xi2022multi}. For ranking metrics, we adopt the widely-used \textit{MAP@K} and \textit{NDCG@K} \cite{ndcg}, consistent with previous work \cite{prm,feng2021revisit,feng2021grn}. For utility metrics, \textit{Click@K}, the expected number of clicks per list, is employed following~\cite{liu2023personalized,xi2024utility}. Formally, \textit{Click@K} is defined as:
\begin{equation}
    \textit{Click@K} = \frac{1}{D} \sum_{i=1}^{D} \left( \sum_{k=1}^{\text{K}} \text{click}_{\omega(k)} \right),
\end{equation}
where $D$ is the total number of re-ranked lists, $\omega$ denotes the ordered list of items, $\omega(k)$ is the item re-ranked at position $k$, and $\text{click}_{\omega(k)}$ denotes whether the item at position $k$ is clicked. For MovieLens-20M whose clicks are generated by DCM~\cite{guo2009efficient}, we continue to adopt DCM to generate clicks on the re-ranked list during evaluation, that is $\text{click}_{\omega(k)}$ are provided by DCM. For the PRM Public dataset with real-world clicks, since we do not know the underlying click model, we employ the common practice of evaluation with the click logs~\cite{xi2022multi,xi2024utility}, meaning $\text{click}_{\omega(k)}$ is whether the item has been clicked in the click logs. For all metrics, we set $K = 5, 10$ for both the MovieLens-20M and PRM Public datasets.

\subsubsection{Reproducibility.}

 We implement our model and baseline methods using Adam~\cite{adam} as the optimizer. For the PRM Public dataset, the maximum list length $M$ is set to 30 for both candidate list and historical list, while for the MovieLens-20M dataset, it is set to 10, adapting to different scenarios. The number of historical lists $N$ is 5 for PRM Public and 3 for the MovieLens-20M dataset. The learning rate is $9\times 10^{-4}$ for PRM Public and $2.5\times 10^{-3}$ for MovieLens-20M dataset. The batch size is 16 for PRM Public and 128 for MovieLens-20M. The hidden size of GRU is 64. The embedding size of categorical features is set to 64, and the MLP of the final prediction layer is defined as [200, 80]. To ensure equitable comparisons, we fine-tune all baseline models to achieve their optimal performance.

\begin{table*}[h]

    \vspace{-10pt}
    \caption{Comparison of RELIFE and its variants. The best result is given in bold, while the second-best value is underlined. }
    \vspace{-5pt}
    \centering
    \scalebox{1}{
    \setlength{\tabcolsep}{1.5mm}
    {\begin{tabular}{cllllll|llllll}
\toprule
\multirow{4}{*}{Model} & \multicolumn{6}{c|}{PRM Public}                                                                                               & \multicolumn{6}{c}{MovieLens-20M}                                                                                                                                                                                                            \\ \cmidrule{2-13} 
                       & \multicolumn{3}{c|}{@5}                                                 & \multicolumn{3}{c|}{@10}                            & \multicolumn{3}{c|}{@5}                                                                                          & \multicolumn{3}{c}{@10}                                                                                        \\ \cmidrule{2-13} 
                       & MAP             & NDCG            & \multicolumn{1}{l|}{Click}        & MAP             & NDCG            & Click         & MAP                                 & NDCG                                & \multicolumn{1}{l|}{Click}         & MAP                                 & NDCG                               & Click                             \\ \midrule
RELIFE-DIM             & 0.3358          & 0.3195          & \multicolumn{1}{l|}{0.6756}         & 0.3513          & 0.3769          & 0.9659          & 0.7513                              & 0.7039                              & \multicolumn{1}{l|}{2.8064}          & 0.7134                              & 0.8080                              & 4.0071                              \\
RELIFE-CPE             & 0.3332          & 0.3184          & \multicolumn{1}{l|}{0.6733}         & 0.3494          & 0.3756          & 0.9627          & 0.7507                              & 0.7035                              & \multicolumn{1}{l|}{2.8053}          & 0.7124                              & 0.8075                             & 4.0072                              \\
RELIFE-SPM             & \underline{0.3388}          & 0.3219          & \multicolumn{1}{l|}{0.6810}          & 0.3538          & 0.3760           & 0.9551          & 0.7563                              & 0.7083                              & \multicolumn{1}{l|}{2.8138}          & 0.7175                              & 0.8112                             & 4.0096                              \\
RELIFE-ICC              & 0.3353          & 0.3162          & \multicolumn{1}{l|}{0.6611}         & 0.3495          & 0.3656          & 0.9132          & 0.7568                              & 0.7102                              & \multicolumn{1}{l|}{2.8265}          & 0.7182                               & 0.8116                            & 4.0040                              \\


RELIFE-CL             & 0.3288          & 0.3154          & \multicolumn{1}{l|}{0.6735}         & 0.3453          & 0.3744          & 0.9759          & 0.7490                               & 0.7022                              & \multicolumn{1}{l|}{2.8041}          & 0.7117                              & 0.8067                             & 4.0055                              \\
RELIFE-PAT             & \underline{0.3388}          & \textbf{0.3236}          & \multicolumn{1}{l|}{\textbf{0.6846}}         & \underline{0.3545}          & \textbf{0.3817}         & \underline{0.9785}          & \underline{0.7578}                               & \underline{0.7115}                              & \multicolumn{1}{l|}{\underline{2.8276}}          & \underline{0.7194}                              & \underline{0.8125}                             & \textbf{4.0158}                             \\ \midrule
\textbf{RELIFE}        & \textbf{0.3393} & \underline{0.3225} & \multicolumn{1}{l|}{\underline{0.6820}} & \textbf{0.3551} & \underline{0.3805} & \textbf{0.9792} & \multicolumn{1}{c}{\textbf{0.7585}} & \multicolumn{1}{c}{\textbf{0.7117}} & \multicolumn{1}{c|}{\textbf{2.8292}} & \multicolumn{1}{c}{\textbf{0.7202}} & \multicolumn{1}{c}{\textbf{0.8130}} & \multicolumn{1}{c}{\underline{4.0122}} \\ \bottomrule
\end{tabular}}
    }

\label{tab:Com}
\vspace{-5pt}
\end{table*}

\vspace{-5pt}
\subsection{Overall Performance (RQ1)}
The overall performance of the two datasets, PRM Public and MovieLens-20M, is presented in Table~\ref{tab:overall}. Several important observations can be inferred from the results. 

First, our proposed RELIFE model consistently outperforms state-of-the-art methods across all metrics on both datasets. As illustrated in Table~\ref{tab:overall}, RELIFE demonstrates superior performance in ranking-based metrics such as \textit{MAP} and \textit{NDCG}, as well as in the utility-based metric \textit{Click}. For instance, on the PRM Public dataset, RELIFE outperforms the strongest baseline, MIR, by 3.50\% in \textit{MAP@10}, 3.88\% in \textit{NDCG@10}, and 4.69\% in \textit{Click@10}. On the MovieLens-20M dataset, RELIFE achieves improvements of 1.24\%, 1.40\%, and 1.06\% over the strongest baseline PIER in \textit{MAP@5}, \textit{NDCG@5}, and \textit{Click@5} respectively. This highlights the effectiveness of incorporating list-level hybrid feedback in modeling user preferences for re-ranking.


Second, most models utilizing users' historical behaviors show better performance compared to models that do not incorporate historical information. Among baselines, MIR and PIER demonstrate superior performance, primarily due to their efficient incorporation of users' history. CTR methods like DFN and RACP improve performance with hybrid feedback; however, without considering multiple candidate items of re-ranking, their performance is generally on par with or inferior to re-ranking methods with positive feedback. These performance gaps highlight the importance of mining user preferences from list-level hybrid feedback and appropriately integrating them to re-ranking. Our proposed RELIFE model achieves the best performance among all models, significantly outperforming the baseline models. The key to this superior performance is that RELIFE incorporates of list-level hybrid feedback and effectively captures a nuanced understanding of user preference and behavior patterns, enhancing re-ranking outcomes.

\vspace{-5pt}
\subsection{In-depth Analysis}
\subsubsection{Ablation Study (RQ2)}\label{sec:ablation}
Several variants of RELIFE are designed to investigate the effectiveness of its components, and we conduct a series of experiments on the PRM Public and MovieLens-20M datasets. Firstly, we explore three variants that remove the main components of RELIFE: \textbf{RELIFE-DIM} excludes the Dual Interest-Miner module, \textbf{RELIFE-CPE} omits the Comparison-aware Pattern Extractor, \textbf{RELIFE-SPM} eliminates the Sequential Preference Mixer, and \textbf{RELIFE-ICC} removes interactions among items within the candidate list, \ie, Intra-Candidate Context modeling. Additionally, we devise two variants to dive into the CPE module: \textbf{RELIFE-CL} removes the contrastive loss used in the CPE while still incorporating intra-list history patterns extracted by CPE in Eq~\eqref{eq:final_mlp}.  \textbf{RELIFE-PAT} omits the intra-list history pattern extraction in Eq~\eqref{eq:final_mlp} but retains the contrastive loss.

The comparison of the aforementioned variants and the original RELIFE on the PRM Public and MovieLens-20M datasets is presented in Table~\ref{tab:Com}. Upon removing each component, there is a noticeable decrease in performance across most metrics, underscoring the importance of each module. Among these variants removing main modules, RELIFE-ICC and RELIFE-CPE exhibit the most significant performance drop, emphasizing the importance of interactions among items within the candidate list and capturing users' comparison behavior. Furthermore, we notice that RELIFE-CL performs worse than RELIFE-CPE. This is somewhat counterintuitive because RELIFE-CPE also removes the contrastive loss, leading us to suspect whether the history pattern extracted by CPE has a positive effect in Eq~\eqref{eq:final_mlp}. To investigate this, we design RELIFE-PAT and observe that its performance declines across most metrics, indicating its usefulness. A possible explanation is that contrastive learning facilitates the model in learning better history patterns, whereas history patterns without alignment may have a negative impact. This highlights the importance of aligning historical and candidate behaviors with contrastive learning.


\subsubsection{Hyper-parameter Study (RQ3)}
\begin{figure}[h]
    \centering
    \includegraphics[width=1\columnwidth]{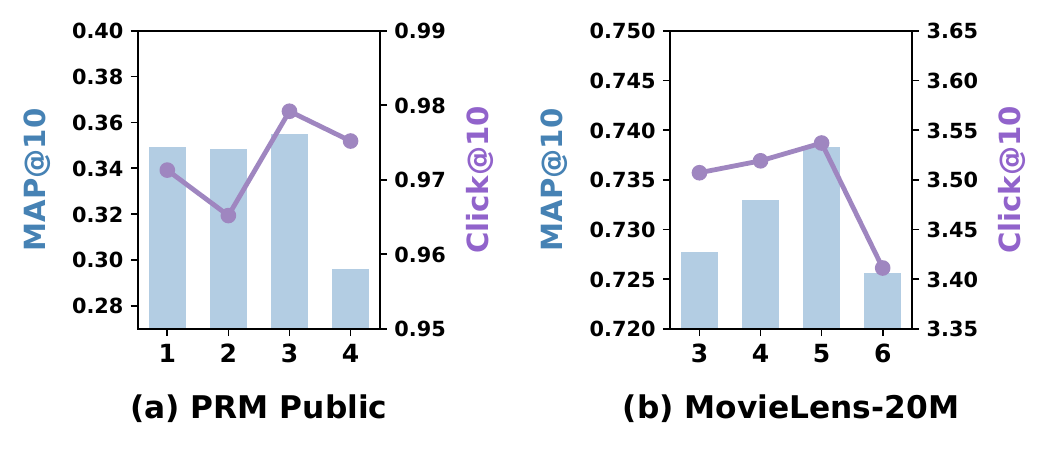}
    \vspace{-24pt}
    \caption{The impact of the number of history lists.}
    \vspace{-10pt}
    \label{fig:number_of_history}
\end{figure}

\begin{figure}[h]
    \centering
    \includegraphics[width=1\columnwidth]{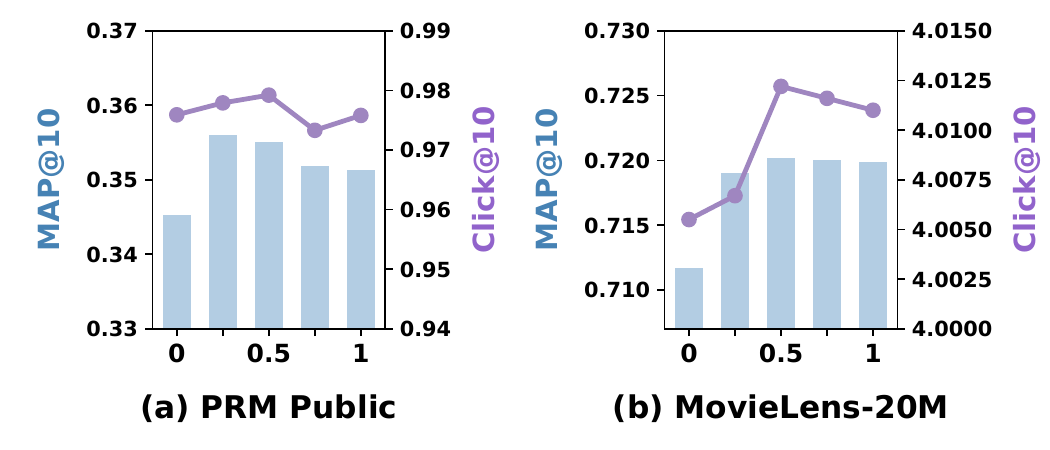}
    \vspace{-24pt}
    \caption{The impact of the weight \(\beta\) of the InfoNCE loss. }
    \vspace{-5pt}
    \label{fig:loss_ratio}
\end{figure}

First, since our focus is on leveraging historical data for re-ranking, the number of history lists serves as a crucial hyperparameter that significantly influences the final results. Therefore, we conduct grid-search experiments on two public datasets to thoroughly explore how varying the number of history lists impacts the performance of RELIFE. Keeping all the other hyperparameters fixed, we adjust the number of history lists and analyze the changes in the ranking-based metric, \textit{MAP@10}, and the utility-based metric, \textit{Click@10}, as illustrated in Figure~\ref{fig:number_of_history}. We observe an increase in both \textit{MAP@10} and \textit{Click@10} from 1 to 3 history lists, followed by stabilization from 3 to 4 on the MovieLens-20M dataset. Similarly, on the PRM Public dataset, \textit{MAP@10} and \textit{Click@10} show an increase from 3 to 5 history lists, followed by a decline from 5 to 6. Based on these observations, we set the number of history lists to 3 for the MovieLens-20M dataset and 5 for the PRM Public dataset in our experiments.

Secondly, based on insights from the ablation study, we acknowledge the crucial role of the Comparison-aware Pattern Extractor, particularly its contrastive learning component. Consequently, we conduct grid-search experiments on public datasets to systematically investigate how the InfoNCE loss impacts the performance of RELIFE. Keeping all other hyperparameters fixed, we vary the weight \(\beta\) of the InfoNCE loss in Eq~\eqref{eq:loss} and analyze its effects on the ranking-based metric, \textit{MAP@10}, and the utility-based metric, \textit{Click@10}, as depicted in Figure~\ref{fig:loss_ratio}. We observe a sharp increase in both \textit{MAP@10} and \textit{Click@10} from 0 to 0.5, followed by stabilization from 0.5 to 1 on the MovieLens-20M dataset. On the PRM Public dataset, \textit{MAP@10} and \textit{Click@10} show improvements from 0 to 0.25, remain relatively stable between 0.25 and 0.5, and decline from 0.5 to 1. Based on these observations, we conclude that the addition of the InfoNCE loss consistently enhances model performance. The performance differences observed with varying values of $\beta$ are relatively small, suggesting that our model is not highly sensitive to this hyperparameter. In our experiments, we set $\beta$ to be $0.5$.



    
    
    
    


\subsubsection{Efficiency Study (RQ4)} To quantify the actual complexity of RELIFE, we compare the inference time and resouece usage of RELIFE with the strongest re-ranking baselines, MIR~\cite{xi2022multi} and PIER~\cite{shi2023pier}, and CTR method DFN~\cite{xie2021deep}.  All the experiments are conducted on the same NVIDIA 3090 GPU of 12G memory with a batch size of 256. Table~\ref{tab:efficiency} presents the experiment results where \textbf{IT} denotes the average inference time for each batch and \textbf{GM} denotes the total GPU memory usage. We can observe RELIFE has slightly higher time and resource consumption than MIR and is on par with PIER and DFN. Considering that RELIFE's improvement is quite significant (about 1-4\%), we believe this overhead is worthwhile. Moreover, RELIFE's inference time can meet the common response latency requirement of industrial RSs, within 100 ms~\cite{xi2023towards}.

\begin{table}[h]
\caption{Inference time and resource usage comparison. }
    \vspace{-6pt}
    \centering
    \scalebox{1}{
    \setlength{\tabcolsep}{2mm}{
\begin{tabular}{ccc|cc}
\toprule
\multirow{2}{*}{Method} & \multicolumn{2}{c|}{PRM Public} & \multicolumn{2}{c}{MovieLens-20M} \\
\cmidrule{2-5}
 & IT (ms) & GM (G) & IT (ms) & GM (G) \\
 \midrule
MIR &  24.54 & 8.57 & 42.03 & 0.44  \\
PIER & 50.08 & 9.38	& 71.84 & 0.69  \\
DFN & 55.98 & 9.37 & 57.82 & 0.91 \\
RELIFE & 61.10 & 10.04 & 60.27 & 1.00 \\
\bottomrule
\end{tabular}
}}
\label{tab:efficiency}
\vspace{-9pt}
\end{table}


\section{Conclusion}
This work highlights the importance of list-level hybrid feedback for re-ranking and proposes RELIFE. For preference mining, we design a Disentangled Interest Miner and a Sequential Preference Mixer to extract preferences from the perspectives of disentanglement and entanglement. Next, we propose a Comparison-aware Pattern Extractor to capture user behavior patterns within each list and leverage contrastive learning to align behavior patterns of candidate and history lists. Extensive experiments show that RELIFE significantly outperforms SOTA baselines. As historical behaviors used in industry are often positive feedback, list-level hybrid feedback shows promising potential for improving re-ranking performance and bringing significant revenue for industrial RSs.

%% file: appendix.tex
\appendix
\section{Dataset}\label{sec:dataset}
Our experiments are conducted on two public datasets, PRM public and MovieLens-20M. 
\begin{itemize}
    \item \textbf{PRM Public} contains 7,246,323 items, 743,720 users, and 14,350,968 records from a real-world e-commerce recommender system. Each record is a recommendation list of 30 items and the user's feedback on them, with 3 user profile features for each user and 5 categorical features for each item. For each user, the last recommendation list in her interacted lists is utilized as the candidate list for re-ranking, and the previous interacted lists are used to construct the history lists. 
    \item \textbf{MovieLens-20M} records 20 million ratings provided by 138, 493 users for 27,278 movies. Although it is not presented in list format, many re-ranking studies utilize this dataset to generate simulated lists for experiments\cite{ren2023slate,liu2023personalized,wu2019pd}. Due to the difficulty in finding datasets that contain both candidate lists and historical behaviors in list form, we follow the approach in re-ranking~\cite{liu2023personalized} and use the ML-20M dataset along with a dependent click model (DCM)~\cite{guo2009efficient,liu2018contextual,katariya2016dcm} to generate simulated candidate lists. First, we sort the users' interaction records by time and divide the user behavior into lists of length 20. Ratings of 4 and 5 are considered positive feedback, while lower ratings are considered negative feedback. Next, We take the last list of user behaviors as the candidate list, with the preceding interaction lists serving as the historical lists. Lastly, following~\cite{liu2023personalized}, we involve a DCM to simulate user clicks on candidate lists for training and evaluation of the re-ranking models on these datasets. Considering multiple clicks and the dependencies between items, DCM can provide unbiased evaluation for a given list and is widely used for click simulation~\cite{liu2018contextual,katariya2016dcm,guo2009efficient}.
\end{itemize}

\section{Baselines}\label{sec:baseline}
We compare the proposed model with the following state-of-the-art re-ranking models, listed as follows
\begin{itemize}
\item \textbf{DLCM}~\cite{dlcm} begins by utilizing a GRU to encode and subsequently re-rank the highest-ranking results.
\item \textbf{PRM}~\cite{prm} employs self-attention to capture the mutual influence between item pairs and user preferences.
\item \textbf{SetRank}~\cite{setrank} creates permutation-equivariant representations for items using self-attention.
\item \textbf{SRGA}~\cite{qian2022scope} uses gated attention mechanisms to effectively capture the interdependence among items within a feed.
\item \textbf{SAR}~\cite{ren2023slate} adopts a slate-aware ranking model that leverages an encoder-decoder architecture to account for item interactions and dependencies, thereby enhancing recommendation relevance.
\item \textbf{PEAR}~\cite{li2022pear} introduces a personalized re-ranking model with a contextualized transformer to capture both feature-level and item-level interactions, considering historical item contexts.
\item \textbf{MIR}~\cite{xi2022multi} proposes a re-ranking framework that integrates user behavior history and utilizes SLAttention to model the interaction between the candidate set and the history list.
\item \textbf{PIER}~\cite{shi2023pier} applies permutation-level interest modeling along with an omnidirectional attention mechanism to improve the efficiency and effectiveness of re-ranking processes in e-commerce.
\end{itemize}

Note that our proposed method is a re-ranking algorithm that optimizes the overall utility of the re-ranking list. Due to differing optimization goals, we do not compare it with algorithms that optimize other objectives, such as diversity~\cite{liu2023personalized,carraro2024enhancing,xu2023multi,wu2019pd} and fairness~\cite{han2023fair,xu2023p}. Additionally, we do not compare with some utility-oriented algorithms, like IRGPR~\cite{irgpr}, because our datasets lack the graph data they need. 

We also compare two works in the CTR domain that utilize both positive and negative feedback. Since they did not address the situation of multiple candidate items in re-ranking, we made the necessary modifications to append the extracted preferences to each candidate item.
\begin{itemize}
    \item \textbf{DFN}~\cite{xie2021deep} jointly consider explicit/implicit and positive/negative feedback to learn user unbiased preferences for recommendation.
    \item \textbf{RACP}~\cite{fan2022modeling} explores the user’s page contextualized behavior sequence with the whole page context for CTR prediction.
\end{itemize}

\begin{figure}[htbp]
    \centering
    \begin{subfigure}[b]{1\columnwidth}
        \centering
        \includegraphics[width=\linewidth]{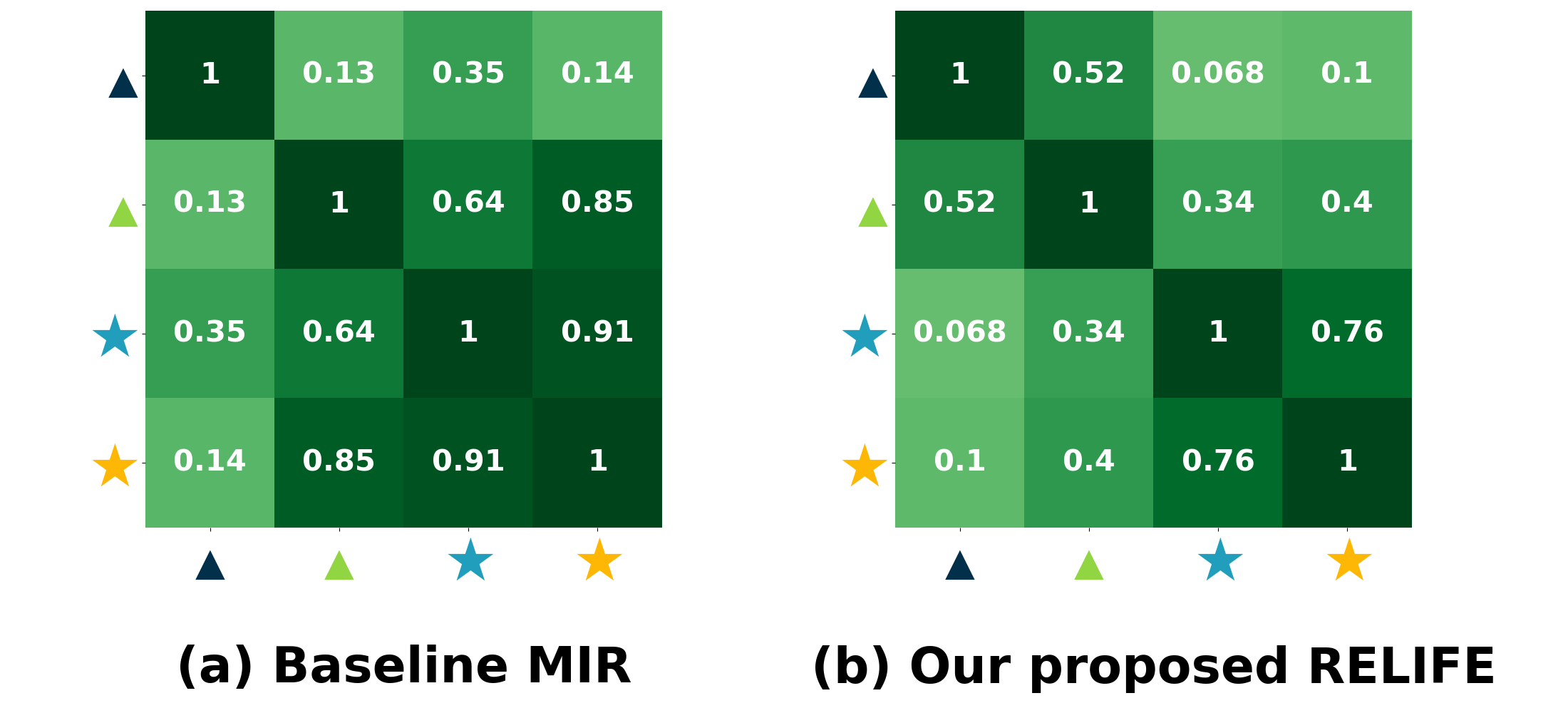}
    \end{subfigure}
    \vspace{-20pt}
    \caption{Similarity between positive and negative item embeddings from MIR and RELIFE. \textcolor[RGB]{2,48,74}{$\blacktriangle$} denotes positive candidate items, \textcolor[RGB]{145,213,66}{$\blacktriangle$} denotes positive historical items. \textcolor[RGB]{33,158,188}{$\bigstar$} denotes negative candidate items, \textcolor[RGB]{254,183,33}{$\bigstar$} denotes negative historical items. 
    }
    \vspace{-15pt}
    \label{fig:casestudy}
\end{figure}

\section{Case Study}
To illustrate the capability of RELIFE in distinguishing between positive and negative items, we visualize the cosine similarity between representations of positive and negative items in a given user's history and candidate lists. Specifically, we select a user from the PRM-Public dataset and obtain item embeddings learned by RELIFE as well as the strongest baseline, MIR, for each item in the user's historical and candidate lists. Subsequently, we compute the average embedding vectors for all positive and negative items in both historical and candidate lists and the cosine similarity between them. The resulting heatmaps for RELIFE and MIR are presented in Figure~\ref{fig:casestudy}, where each cell represents the cosine similarity between item embeddings.


From Figure~\ref{fig:casestudy}, it is evident that RELIFE is more adept than MIR at learning representations that effectively differentiate between positive and negative items. In Figure~\ref{fig:casestudy} (b), the positive candidate item is more similar to the positive history item, while their similarity to negative items is relatively low, indicating a high degree of distinction between positive and negative items. Conversely, in Figure~\ref{fig:casestudy} (a), the similarity among positive items is sometimes lower than that between positive and negative items, suggesting an inability to distinguish them effectively. This demonstrates that RELIFE, which incorporates list-level history, can learn superior representations for distinguishing positive and negative items.